\documentclass[aps,twocolumn,amsmath]{revtex4}
\usepackage{graphicx}
\usepackage{dcolumn}
\def\Journal#1#2#3#4{{#1} {\bf#2}, #3 (#4)}

\def\NPA{{\rm Nucl. Phys.} A}

\def\PLB{{\rm Phys. Lett.}  B}
\def\PRL{\rm Phys. Rev. Lett.}
\def\PRD{{\rm Phys. Rev.} D}

\def\JPG{{\rm J. Phys.} G}

\def\ep{\epsilon}
\def\lam{\lambda}

\def\la{\langle}
\def\ra{\rangle}

\def\al{\alpha}

\def\be{\begin{equation}}
\def\ee{\end{equation}}
\def\bea{\begin{eqnarray}}
\def\eea{\end{eqnarray}}

\begin{document}
\title{Light-front quark model analysis of heavy meson radiative decays}
\author{ Ho-Meoyng Choi}
\affiliation{ Department of Physics, Teachers College, 
Kyungpook National University, Daegu, Korea 702-701}
\begin{abstract}
We present the magnetic dipole($M1$) transitions $V\to P\gamma$ 
of various heavy-flavored mesons such as $(D,D^*,D_s,D^{*}_s,\eta_c, J/\psi)$
and $(B,B^*,B_s,B^*_s,\eta_b,\Upsilon)$ using the light-front quark model 
constrained by the variational principle for the QCD-motivated effective
Hamiltonian. The weak decay constants of heavy mesons and the 
decay widths for $V\to P\gamma$ are calculated. 
The radiative decay for $\Upsilon\to\eta_{b}\gamma$ process is found to be
very helpful to determine the unmeasured mass of $\eta_b$. 
Our numerical results are overall in good 
agreement with the available experimental data as well as other 
theoretical model calculations. 
\end{abstract}


\maketitle
\section{Introduction}
The physics of exclusive heavy meson decays
has provided very useful testing ground for the precise 
determination of the fundamental parameters of the standard model(SM) and 
the development of a better understanding of the QCD dynamics.
While the experimental tests of exclusive heavy meson decays are 
much easier than those of inclusive one, the theoretical understanding 
of exclusive decays is complicated mainly due to the nonperturbative
hadronic matrix elements entered in the long distance nonperturbative
contributions. Since a rigrous field-theoretic formulation
with a first principle application of QCD to make a 
reliable estimates of the nonperturbative hadronic matrix elements
has not so far been possible, most of theoretical efforts have been 
devoted to looking for phenomenological approaches to nonperturbative 
QCD dynamics.

In our previous light-front quark model(LFQM) analysis~\cite{CJ1}
based on the QCD-motivated effective Hamiltonian, we have analyzed various 
exlcusive processes such as the semileptonic decays between heavy 
pseudoscalar mesons~\cite{CJ2} and the rare $B\to K$ 
decays~\cite{CJ3} and found a good agreement with the experimental data.
Along with those exclusive processes, the magnetic dipole($M1$) transitions 
$V(1^3S_1)\to P(1^1S_0)\gamma$ from the spin-triplet $S$-wave vector(V) 
mesons to the spin-singlet $S$-wave pseudoscalar(P) mesons have also been 
considered as a valuable testing ground to further constrain the 
phenomenological model of 
hadrons~\cite{GI,Barik,Jaus96,EFG1, Cheng,HQET, CBM, Amu, Nora, EFG2}. 

In this talk we thus investigate the magnetic dipole transition among
the heavy-flavored mesons such as $(D,D^*,D_s,D^{*}_s,\eta_c, J/\psi)$ and
$(B,B^*,B_s,B^*_s,\eta_b,\Upsilon)$ using our LFQM~\cite{CJ1,CJ2,CJ3}. 
Since the experimental data available in this heavy-flavored sector are scanty, 
predictions of a model, if found reliable, can be utilized quite
fruitfully. In addition, we calculate the weak decay constants of heavy
pseudoscalar and vector mesons. 
A reliable estimate of decay constants is important, as they appear in
may processes from which we can extract fundamental quantities in the SM
such as Cabibbo-Kobayashi-Maskawa matrix elements. 
In our LFQM~\cite{CJ1,CJ2,CJ3}, we have implemented the variational 
principle to QCD-motivated effective LF Hamiltonian to enable us to 
analyze the meson mass spectra and to find optimized model parameters, 
which are to be used subsequently in the present investigation. 
Such an approach can better constrain the phenomelogical parameters and  
establish the extent of applicability of our LFQM to wider ranging 
hadronic phenomena.

The paper is organized as follows: 
In Sec.II, we briefly describe the formulation of our LFQM~\cite{CJ1,CJ2}
and the procedure of fixing the model parameters using the variational
principle for the QCD-motivated effective Hamiltonian. The decay constants
and radiative $V\to P\gamma$ decay widths for heavy-flavored mesons are
then uniquely determined in our model calculation.
In Sec. III, the formulae for the decay constants of pseudoscalar
and vector mesons as well as the decay widths for $V\to P\gamma$ in our
LFQM are given. To obtain the $q^2$-dependent transition form factors
$F_{VP}(q^2)$ for $V\to P\gamma^*$ transitions, we use the 
Drell-Yan-West $q^+=0$ frame(i.e.
$q^2=-{\bf q}^2_\perp<0$) and then analytically continue the spacelike
results to the timelike $q^2>0$ region by changing ${\bf q}_\perp$ to 
$i{\bf q}_\perp$ in the form factor. The coupling constants
$g_{VP\gamma}$ needed for the calculations of the decay widths for
$V\to P\gamma$ can then be determined in the limit as 
$q^2\to 0$, i.e. $g_{VP\gamma}=F_{VP}(q^2=0)$.
In Sec. IV, we present our numerical results and compare
with the available experimental data as well as other theoretical 
model predictions. Summary and conclusions follow in Sec.V.
 
\section{Model Description}
The key idea in our LFQM~\cite{CJ1,CJ2} for mesons is to treat the radial 
wave function as trial function for the variational principle to the 
QCD-motivated effective Hamiltonian saturating the Fock state expansion 
by the constituent quark and antiquark. The QCD-motivated Hamiltonian for 
a description of the ground state meson mass spectra is given by
\bea\label{Ham}
H_{q\bar{q}}|\Psi^{JJ_z}_{nlm}\ra&=&\biggl[
\sqrt{m^2_q+{\vec k}^2}+\sqrt{m^2_{\bar{q}}+{\vec k}^2}+V_{q\bar{q}}\biggr]
|\Psi^{JJ_z}_{nlm}\ra,
\nonumber\\
&=&[H_0 + V_{q\bar{q}}]|\Psi^{JJ_z}_{nlm}\ra
=M_{q\bar{q}}|\Psi^{JJ_z}_{nlm}\ra,
\eea
where ${\vec k}=({\bf k}_\perp, k_z)$ is the three-momentum of the 
constituent quark, $M_{q\bar{q}}$ is the mass 
of the meson, and $|\Psi^{JJ_z}_{nlm}\ra$ is the meson
wave function. In this work, we use two interaction potentials $V_{q\bar{q}}$
for the pseudoscalar($0^{-+}$) and vector($1^{--}$) mesons: (1) Coulomb
plus harmonic oscllator(HO), and (2) Coulomb plus linear confining potentials.
In addition, the hyperfine interaction, which is essential to distinguish
vector from pseudoscalar mesons, is included for both cases, viz.,
\be\label{pot}
V_{q\bar{q}}=V_0 + V_{\rm hyp}
= a + {\cal V}_{\rm conf}-\frac{4\al_s}{3r}
+\frac{2}{3}\frac{{\bf S}_q\cdot{\bf S}_{\bar{q}}}{m_qm_{\bar{q}}}
\nabla^2V_{\rm coul},
\ee
where ${\cal V}_{\rm conf}=br(r^2)$ for the linear(HO) potential and
$\la{\bf S}_q\cdot{\bf S}_{\bar{q}}\ra=1/4(-3/4)$ for the 
vector(pseudoscalar) meson.

The momentum space light-front wave function of the ground state
pseudoscalar and vector mesons is given by 
\be\label{w.f}
\Psi^{JJ_z}_{100}(x_i,{\bf k}_{i\perp},\lam_i)
={\cal R}^{JJ_z}_{\lam_1\lam_2}(x_i,{\bf k}_{i\perp})
\phi(x_i,{\bf k}_{i\perp}),
\ee
where $\phi(x_i,{\bf k}_{i\perp})$ is the radial wave function and 
${\cal R}^{JJ_z}_{\lam_1\lam_2}$ is the spin-orbit wave function,
which is obtained by the interaction independent Melosh transformation
from the ordinary equal-time static spin-orbit wave function assigned
by the quantum numbers $J^{PC}$.
The model wave function in Eq.~(\ref{w.f}) is represented by the
Lorentz-invariant variables, $x_i=p^+_i/P^+$, 
${\bf k}_{i\perp}={\bf p}_{i\perp}-x_i{\bf P}_\perp$ and $\lam_i$, where
$P^\mu=(P^+,P^-,{\bf P}_\perp)
=(P^0+P^3,(M^2+{\bf P}^2_\perp)/P^+,{\bf P}_\perp)$ is the momentum of the
meson $M$, $p^\mu_i$ and $\lam_i$ are the momenta and the helicities of 
constituent quarks, respectively.

The covariant forms of the spin-orbit wave functions
for pseudoscalar and vector mesons are given by 
\bea\label{R00_A}
{\cal R}_{\lam_1\lam_2}^{00}
&=&\frac{-\bar{u}(p_1,\lam_1)\gamma_5 v(p_2,\lam_2)}
{\sqrt{2}\tilde{M_0}},
\nonumber\\
{\cal R}_{\lam_1\lam_2}^{1J_z}
&=&\frac{-\bar{u}(p_1,\lam_1)
\biggl[/\!\!\!\ep(J_z) -\frac{\ep\cdot(p_1-p_2)}{M_0 + m_1 + m_2}\biggr]
v(p_2,\lam_2)} {\sqrt{2}\tilde{M_0}},
\nonumber\\
\eea
where $\ep^\mu(J_z)$ is the polarization vector
of the vector meson~\cite{DA}, $\tilde{M_0}=\sqrt{M^2_0-(m_1-m_2)^2}$ 
and $M^2_0$  is the invariant meson mass square $M^2_0$ defined as 
$M^2_0=\sum_{i=1}^2\frac{{\bf k}^2_{i\perp}+m^2_i}{x_i}$.
The spin-orbit wave functions satisfy the following relations
$\sum_{\lam_1\lam_2}{\cal R}_{\lam_1\lam_2}^{JJ_z\dagger}
{\cal R}_{\lam_1\lam_2}^{JJ_z}=1$,
for both pseudoscalar and vector mesons.
For the radial wave function $\phi$, we use the same Gaussian wave function 
for both pseudoscalar and vector mesons 
\be\label{rad}
\phi(x_i,{\bf k}_{i\perp})=\frac{4\pi^{3/4}}{\beta^{3/2}}
\sqrt{\frac{\partial k_z}{\partial x}}
{\rm exp}(-{\vec k}^2/2\beta^2),
\ee
where $\beta$ is the variational parameter.
When the longitudinal component $k_z$ is defined by 
$k_z=(x-1/2)M_0 + (m^2_2-m^2_1)/2M_0$, the Jacobian of the variable
transformation $\{x,{\bf k}_\perp\}\to {\vec k}=({\bf k}_\perp, k_z)$
is given by
\be\label{jacob}
\frac{\partial k_z}{\partial x}=\frac{M_0}{4x_1x_2}
\biggl\{ 1- \biggl[\frac{m^2_1-m^2_2}{M^2_0}\biggr]^2\biggr\}.
\ee
The normalization factor in Eq.~(\ref{rad}) is obtained from the 
following normalization of the total wave function,
\be\label{norm}
\int^1_0dx\int\frac{d^2{\bf k}_\perp}{16\pi^3}
|\Psi^{JJ_z}_{100}(x,{\bf k}_{i\perp})|^2=1.
\ee
Our variational principle to the QCD-motivated effective Hamiltonian
first evaluate the expectation value of the central Hamiltonian $H_0+V_0$,
i.e. $\la\phi|(H_0+V_0)|\phi\ra$ with a
trial function $\phi(x_i,{\bf k}_{i\perp})$ that depends on the
variational parameters $\beta$ and varies $\beta$ until 
$\la\phi|(H_0+V_0)|\phi\ra$ is a minimum.  Once these model
parameters are fixed, then, the mass eigenvalue of each meson is obtained
by $M_{q\bar{q}}=\la\phi|(H_0+V_{q\bar{q}})|\phi\ra$.
More detailed procedure of determining the model parameters of light and
heavy quark sectors can be found in our previous works~\cite{CJ1,CJ2}.
Our model parameters $(m,\beta)$ for the heavy quark sector obtained 
from the linear and HO potential models are summarized in Table~\ref{t1}.
\begin{table*}[t]
\caption{The constituent quark mass[GeV] and the Gaussian paramters
$\beta$[GeV] for the linear and HO potentials obtained by the variational
principle. $q=u$ and $d$.}\label{t1}
\begin{tabular}{ccccccccccc} \hline\hline
Model & $m_q$ & $m_s$ & $m_c$ & $m_b$ & $\beta_{qc}$
& $\beta_{sc}$ & $\beta_{cc}$ & $\beta_{qb}$ & $\beta_{sb}$
& $\beta_{bb}$ \\
\hline
Linear & 0.22 & 0.45 & 1.8 & 5.2 &
0.468 & 0.502 & 0.651 & 0.527 & 0.571 & 1.145\\
\hline
HO & 0.25 & 0.48 & 1.8 & 5.2 &
0.422 & 0.469 & 0.700 & 0.496 & 0.574 & 1.803\\
\hline\hline
\end{tabular}
\end{table*}

\begin{figure}
\vspace{1.2cm}
\includegraphics[width=3in,height=3in]{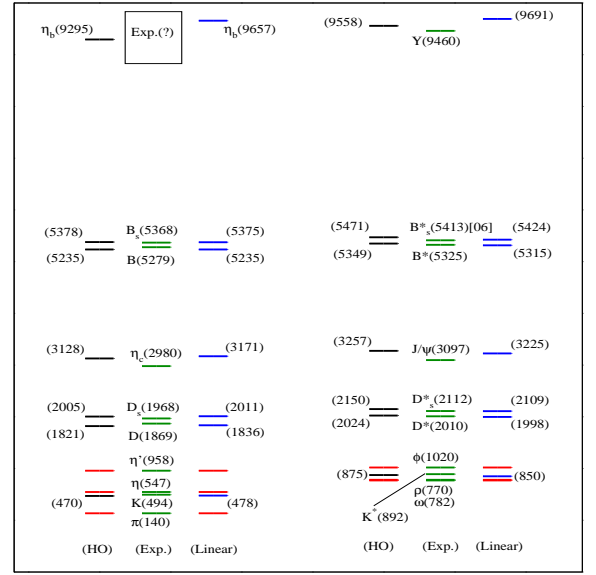}
\caption{\label{fig1}(Color online).
Fit of the ground state meson masses[MeV] with the parameters given
in Table~\ref{t1}. The $(\rho,\pi)$, $(\eta,\eta')$, and $(\omega,\phi)$ 
masses are our input data(red online).}
\end{figure}

Our predictions of the ground state meson mass spectra obtained from 
the linear and HO potential parameters are summarized in Fig.~\ref{fig1}.
As one can see, our predictions obtained from both
linear and HO parameters are overall in good agreement with the 
data~\cite{Data06} within 6$\%$ error. 
As we shall see in our numerical calculations, 
the radiative decay of $\Upsilon\to\eta_b\gamma$ might be useful to 
determine the mass of $\eta_b$ experimentally 
since the decay width $\Gamma(\Upsilon\to\eta_b\gamma)$ is 
very sensitive to the value of $\Delta m(=M_\Upsilon-M_{\eta_b})$, viz. 
$\Gamma\propto(\Delta m)^3$.

The decay constants of pseudoscalar and vector mesons are defined by 
\bea\label{fp}
\la 0|\bar{q}\gamma^\mu\gamma_5 q|P\ra&=&if_P P^\mu,
\nonumber\\
\la 0|\bar{q}\gamma^\mu q|V(P,h)\ra&=&f_V M_V\ep^\mu(h),
\eea
where the experimental value of vector meson decay constant $f_V$ 
is extracted from the longitudinal($h=0$) polarization. 
Using the plus component($\mu=+$) of the current, one can easily calculate the
decay constants and the explicit forms of pseudoscalar and vector meson
decay constants are given in~\cite{DA}.

\section{Radiative decay width for $V\to P\gamma$ }
\begin{figure}[b]
\vspace{1cm}
\includegraphics[width=3.0in,height=1.5in]{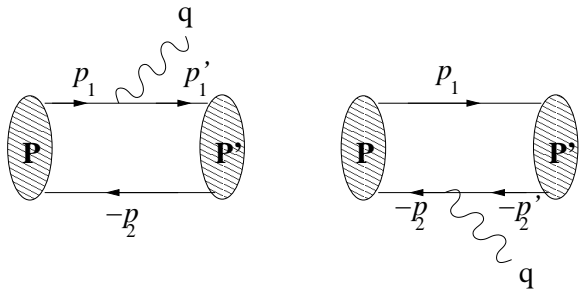}
\caption{\label{fig2}Lowes-order graph for $V\to P\gamma^*$ transitions.} 
\end{figure}

In our LFQM calculation of $V\to P\gamma$ decay process, we shall first 
analyze the virtual photon($\gamma^*$) decay process so that we calculate
the momentum dependent transition form factor, $F_{VP}(q^2)$. 
The lowest-order Feynman diagram for $V\to P\gamma^*$ process is
shown in Fig.~\ref{fig2} where the decay from vector meson to pseudoscalar
meson and virtual photon state is mediated by a quark loop with flavors
of consituent mass $m_1$ and $m_2$.

The transition form factor $F_{VP}(q^2)$ for the magnetic dipole 
decay of vector meson $V(P)\to P(P')\gamma^*(q)$ is defined as \be\label{ff}
\la P(P')|J^\mu_{\rm em}|V(P,h)\ra
=ie\epsilon^{\mu\nu\rho\sigma}\epsilon_\nu(P,h)
q_{\rho}P_{\sigma}F_{VP}(q^2),
\ee
where $q=P-P'$ is the four
momentum of the virtual photon, $\ep_\nu(P,h)$ is the polarization 
vector of the initial meson with four momentum $P$ and helicity $h$. 
The kinematically allowed momentum transfer squared $q^2$ ranges 
from 0 to $q^2_{\rm max}=(M_V-M_P)^2$.

The decay form factor $F_{VP}(q^2)$ can be obtained in the 
$q^+=0$ frame with the ``good" component of currents, i.e.
$\mu=+$, without encountering zero-mode contributions~\cite{ZM}.
Thus, we shall perform our LFQM calculation in the $q^+=0$ frame, where
$q^2=q^+q^- - {\bf q}^2_\perp=-{\bf q}^2_\perp<0$, and then analytically
continue the form factor $F_{VP}({\bf q}^2_\perp)$ in the spacelike region
to the timelike $q^2>0$ region by changing ${\bf q}_\perp$ to 
$i{\bf q}_\perp$ in the form factor.
In the calculations of the decay form factor $F_{VP}(q^2)$, we use
`+'-component of currents and the transverse($h=\pm 1$) polarization.

The hadronic matrix element of the plus current, 
$\la J^+\ra\equiv \la P(P')|J^+_{\rm em}|V(P,h=+)\ra$ in
Eq.~(\ref{ff}) is then obtained by the convolution formula of the initial 
and final state light-front wave functions: 
\bea\label{Jplus}
\la J^+\ra&=&\sum_{j}ee_j\int^1_0\frac{dx}{16\pi^3}
\int d^2{\bf k}_\perp\phi(x,{\bf k'}_\perp)
\phi(x,{\bf k}_\perp)
\nonumber\\
&&\times\sum_{\lam\bar{\lam}}{\cal R}^{00\dagger}_{\lam'\bar{\lam}}
\frac{\bar{u}_{\lam'}(p'_1)}{\sqrt{p'^+_1}}\gamma^+
\frac{u_{\lam}(p_1)}{\sqrt{p^+_1}}
{\cal R}^{11}_{\lam\bar{\lam}},
\eea
where ${\bf k'}_\perp={\bf k}_\perp - x_2{\bf q}_\perp$
and $ee_j$ is the electrical charge for $j$-th quark flavor. 
Comparing with the right-hand-side of Eq.~(\ref{ff}), i.e.
$eP^+F_{VP}(Q^2)q^R/\sqrt{2}$ where $q^R=q_x +iq_y$, 
we could extract the one-loop integral, $I(m_1,m_2,q^2)$,
which is given by 
\bea\label{soft_form}
I(m_1,m_2,q^2) &=&\int^1_0 \frac{dx}{8\pi^3}\int d^2{\bf k}_\perp
\frac{\phi(x, {\bf k'}_\perp)\phi(x,{\bf k}_\perp)}
{x_1\tilde{M_0}\tilde{M'_0}}
\nonumber\\
&&\times
\biggl\{{\cal A} 
+ \frac{2}
{{\cal M}_0}
[{\bf k}^2_\perp
-
\frac{({\bf k}_\perp\cdot{\bf q}_\perp)^2}{{\bf q}^2_\perp}]
\biggr\},
\nonumber\\
\eea
where the primed factors are the functions of final state momenta,
e.g. $\tilde{M_0}'=\tilde{M_0}'(x,{\bf k}'_\perp)$. 

Then, the decay form factor $F_{VP}(q^2)$ is obtained as
\be\label{FS}
F_{VP}(q^2)= e_1I(m_1,m_2,q^2) + e_2 I(m_2,m_1,q^2).
\ee
The coupling constant $g_{VP\gamma}$ for real photon($\gamma$) case
can then be determined in the limit
as $q^2\to 0$, i.e. $g_{VP\gamma}=F_{VP}(q^2=0)$.
The decay width for $V\to P\gamma$ is given by 
\be\label{width}
\Gamma(V\to P\gamma)=\frac{\alpha}{3}g_{VP\gamma}^2 k^3_\gamma,
\ee
where $\alpha$ is the fine-structure constant  and
$k_\gamma=(M^2_V-M^2_P)/2M_V$ is the kinematically allowed energy
of the outgoing photon.

\section{Numerical Results}
In our numerical calculations, we use two sets of model parameters
($m,\beta$) for the linear and HO confining potentials 
given in Table~\ref{t1}
to perform, in a way, a parameter-free-calculation of decay constants 
and decay rates for heavy pseudoscalar and vector mesons.
Although our predictions of ground state heavy meson masses are overall 
in good agreement with the experimental values,
we use the experimental meson masses except $\eta_b$ meson in the 
computations of the radiative decay widths to reduce possible theoretical 
uncertainties. Since the $\eta_b$ mass is not measured yet, 
we use the range $\Delta m = M_{\Upsilon}-M_{\eta_b}=60\sim160$ MeV 
for $\Upsilon\to\eta_b\gamma$ process~\cite{ALEPH}.

\begin{table*}[t]
\caption{Charmed meson decay constants(in unit of MeV) obtained
from the linear[HO] parameters.}\label{t2}
\begin{tabular}{ccccccc} \hline\hline
& $f_D$ & $f_{D^*}$ & $f_{D_s}$ & $f_{D^*_s}$ &$f_{\eta_c}$ 
& $f_{J/\psi}$  \\
\hline
Linear[HO] & 211[194] & 254[228] & 248[233] & 290[268] 
& 326[354] & 360[395] \\
\hline
Lattice~\cite{Bec}   & $211\pm14^{+2}_{-12}$
& $245\pm20^{+3}_{-2}$ & $231\pm12^{+8}_{-1}$ 
& $272\pm16^{+3}_{-20}$ & $-$ & $-$ \\
QCD\;\;\;~\cite{Aubin}   & $201\pm3\pm17$
& $-$ & $249\pm3\pm16$ & - & $-$ & $-$ \\
\hline
Sum-rules~\cite{Nar}   & $204\pm20$
& $-$ & $235\pm24$ & - & $-$ & $-$ \\
\hline
BS~\cite{Wang}   & $230\pm25$ & $340\pm23$ & $248\pm27$ 
& $375\pm24$& $292\pm25$ & $459\pm28$ \\ 
\hline
QM~\cite{CG} & $240\pm20$ & $-$ & $290\pm 20$ & $-$ & $-$ & $-$ \\
RQM~\cite{Ebert} & 234 & 310 & 268 & 315 
&$-$ & $-$ \\
\hline
Exp.        & $222.6\pm16.7^{+2.8}_{-3.4}$~\cite{Cleo05} 
& $-$ & $274\pm13\pm7$~\cite{Cleo07} &-& $335\pm 75$~\cite{Cleo_eta} 
& $416\pm6$~\cite{Data06} \\
\hline\hline
\end{tabular}
\end{table*}
In Tables~\ref{t2} and~\ref{t3}, we present our predictions for 
the charmed and bottomed meson decay constants, respectively,
and compare them with other theoretical model 
predictions~\cite{Bec,Aubin,Nar,Wang,CG,Ebert,Gray,Hash,Jamin} as well as 
the experimental data~\cite{Data06,Cleo05,Cleo07,Cleo_eta,Belle_B}. 
Our predictions for the ratios 
$f_{D_s}/f_D=1.18[1.20]$ and $f_{\eta_c}/f_{J/\psi}=0.91[0.90]$
obtained from the linear[HO] parameters are in good agreement 
with the available experimental data, 
$(f_{D_s}/f_D)_{\rm exp.}=1.23\pm0.11\pm0.04$~\cite{Cleo07}
and $(f_{\eta_c}/f_{J/\psi})_{\rm exp.}=0.81\pm0.19$~\cite{Cleo_eta,Data06},
respectively. Our results for the ratios $f_{B_s}/f_{B}=1.24[1.32]$
and $f_{B^*_s}/f_{B^*}=1.23[1.32]$ obtained from the
linear[HO] parameters are quite comparable with the recent lattice results,
$1.20(3)(1)$~\cite{Gray} and $1.22(^{+5}_{-6})$~\cite{Hash}
for $f_{B^*_s}/f_{B^*}$ and
$1.17(4)^{+1}_{-3}$~\cite{Bec} for $f_{B^*_s}/f_{B^*}$.

\begin{table*}[t]
\caption{Bottomed meson decay constants(in unit of MeV) 
obtained from the linear[HO] parameters.}\label{t3}
\begin{tabular}{ccccccc} \hline\hline
 & $f_B$ & $f_{B^*}$ & $f_{B_s}$ & $f_{B^*_s}$ & $f_{\eta_b}$
& $f_{\Upsilon}$  \\
\hline
Linear[HO] & 189[180] & 204[193] & 234[237] & 250[254] 
& 507[897] & 529[983] \\
\hline
Lattice~\cite{Bec}  & $179\pm18^{+34}_{-9}$
& $196\pm24^{+39}_{-2}$ & $204\pm16^{+36}_{-0}$
& $229\pm20^{+41}_{-16}$ & $-$ & $-$ \\
QCD\hspace{0.2cm}~\cite{Gray}  & $216\pm22$ & $-$ & $259\pm32$ 
& $-$ & $-$ & $-$ \\
\hspace{0.9cm}~\cite{Hash}& $189\pm27$ & $-$ & $230\pm30$ & - & $-$ & $-$ \\
\hline
Sum-rules~\cite{Jamin}    & $210\pm19$
& $-$ & $244\pm21$ & - & $-$ & $-$ \\
\hspace{0.9cm}~\cite{Nar}   & $203\pm23$
& $-$ & $236\pm30$ & - & $-$ & $-$ \\
\hline
BS~\cite{Wang}   & $196\pm29$ & $238\pm18$ & $216\pm 32$ & $272\pm20$ 
&$-$ & $498\pm20$ \\
\hline
QM~\cite{CG} & $155\pm 15$ & $-$ & $210\pm 20$ & $-$ & $-$ & $-$ \\
RQM~\cite{Ebert} & 189 & 219 & 218 & 251 
&$-$ & $-$ \\
\hline
Exp.    & $229^{+36+34}_{-31-37}$~\cite{Belle_B} 
& $-$ & $-$ & - & $-$ & $715\pm5$~\cite{Data06} \\
\hline\hline
\end{tabular}
\end{table*}

\begin{figure}[t]
\vspace{1cm}
\includegraphics[width=3.0in,height=3.0in]{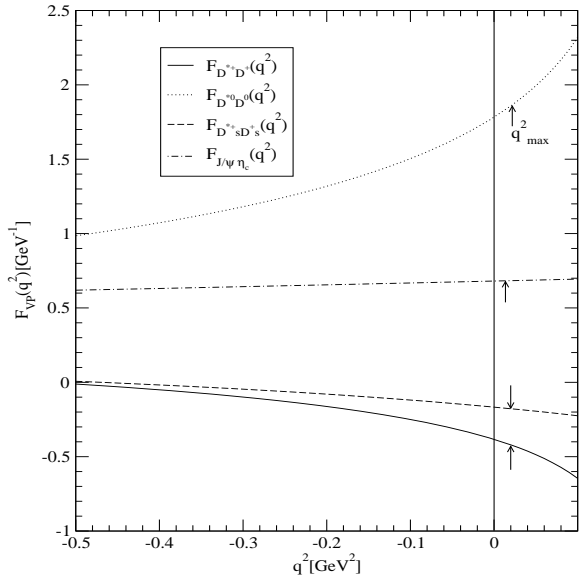}
\caption{\label{fig3}Transition form factors $F_{VP}(q^2)$ for charmed
mesons radiative decays obtained from the linear parameters.}
\end{figure}
We show in Fig.~\ref{fig3} the momentum dependent form factors
$F_{VP}(q^2)$ for charmed vector meson radiative $V\to P\gamma^*$ decays
obtained from the linear parameters.
Since the results from the HO parameters are not much different
from those of linear ones, we omit them for simplicity. 
The arrows in the figure represent the zero recoil points
of the final state pseudoscalar meson, i.e.  $q^2=q^2_{\rm max}$.
We have performed the analytical continuation of the decay form
factors $F_{VP}(q^2)$ from the spacelike region($q^2<0$) to the physical
timelike region $0\leq q^2\leq q^2_{\rm max}$.
The coupling constant $g_{VP\gamma}$ at $q^2=0$ corresponds
to a final state pseudoscalar meson recoiling with maximum
three-momentum in the rest frame
of vector meson. The opposite sign of coupling constants for $D^{*+}$(solid
line) and $D^{*+}_s$(dashed line) decays compared to the 
charmonium $J/\psi$(dot-dashed line) decay indicates
that the charmed quark contribution is largely destructive in the
radiative decays of $D^{*+}$ and $D^{*+}_s$ mesons.
The recoil effect, i.e. the difference between the
zero and the maximum points, is not 
negligible for the $D^{*+}\to D^+\gamma^*$ decay, while other processes
may be negligible.
The recoil effects for the bottomed and bottomonium meson decays are 
negligible due to the very small photon energies.

\begin{table*}[t]
\caption{Decay widths and branching ratios 
for radiative $V\to P\gamma$ decays obtained from our
linear[HO] model parameters. 
We used $M_{\eta_b}=9353\pm50$ MeV for $\Upsilon\to\eta_b\gamma$
decay.}\label{t5}
\begin{tabular}{cccc} \hline\hline
Decay mode & $\Gamma$[keV] & 
$\rm Br$ &  Br$_{\rm exp}$~\cite{Data06} \\
\hline
$J/\psi\to\eta_c\gamma$ 
& $1.69\pm0.05[1.65\pm0.05]$ 
&$(1.80\pm 0.10)[1.76\pm 0.10]\%$ & $(1.3\pm 0.4)\%$ \\
\hline
$D^{*+}\to D^+\gamma$ & $0.90\pm0.02[0.96\pm0.02]$ 
&$(0.93\pm0.31)[1.00\pm 0.34]\%$ & $(1.6\pm 0.4)\%$ \\
\hline
$D^{*0}\to D^0\gamma$ & $20.0\pm0.3[21.0\pm0.3]$
&- & $(38.1\pm 2.9)\%$ \\
\hline
$D^{*+}_s\to D^+_s\gamma$ & $0.18\pm0.01[0.17\pm0.01]$
&- & $(94.2\pm 0.7)\%$ \\
\hline
$B^{*+}\to B^+\gamma$ & $0.40\pm0.03[0.40\pm0.03]$
&- & $-$ \\
\hline
$B^{*0}\to B^0\gamma$ & $0.13\pm0.01[0.13\pm0.01]$
&- & $-$ \\
\hline
$B^{*0}_s\to B^0_s\gamma$ 
& $0.068\pm0.017[0.064\pm0.016]$
&- & $-$ \\
\hline
$\Upsilon\to\eta_b\gamma$ 
& $0.045^{+0.097}_{-0.038}[0.042^{+0.088}_{-0.036}]$ 
& $(8.4^{+18.6}_{-7.2})[7.7^{+17.0}_{-6.6}]\times10^{-4}$ & $-$ \\
\hline\hline
\end{tabular}
\end{table*}

In Table~\ref{t5}, we present our results for the decay  widths and 
branching ratios together with the available experimental data. 
The errors in our results for the decay widths and branching ratios
come from the uncertainties of the experimental mass 
values and experimental mass values plus the full widths, respectively. 
Our results of the branching ratios
${\rm Br}(J/\psi\to\eta_c\gamma)=1.80\pm 0.10[1.76\pm 0.10]\%$
and
${\rm Br}(D^*\to D^+\gamma)= 0.93\pm0.31[1.00\pm0.34]\%$
obtained from the linear[HO] parameters are in agreement with the
experimental data~\cite{Data06}, 
${\rm Br}(J/\psi\to\eta_c\gamma)_{\rm exp}=(1.3\pm 0.4)\%$
and
${\rm Br}(D^*\to D^+\gamma)_{\rm exp}=(1.6\pm 0.4)\%$ within the error bars. 
For the $\Upsilon\to\eta_b\gamma$ process, our predictions for the
decay width and branching ratio obtained
from the linear[HO] parameters are 
$\Gamma(\Upsilon\to\eta_b\gamma)
=45^{+97}_{-38}[42^{+88}_{-36}]\;{\rm eV}$ and
${\rm Br}(\Upsilon\to\eta_b\gamma)
=(8.4^{+18.6}_{-7.2})[7.7^{+17.0}_{-6.6}]\times 10^{-4}$,
where the lower, central, and upper values correspond to $\Delta m=60$ MeV,
110 MeV, and 160 MeV, respectively. 
The decay width $\Gamma(\Upsilon\to\eta_b\gamma)$ is found to be very
sensitive to $\Delta m$ because it is proportional to $(\Delta m)^3$.
Other model calculations for the
$\Upsilon(1S)$ radiative $M1$ decay rates can be found in Ref.~\cite{GR}. 

\begin{figure}[t]
\vspace{1cm}
\includegraphics[width=3.0in,height=3.0in]{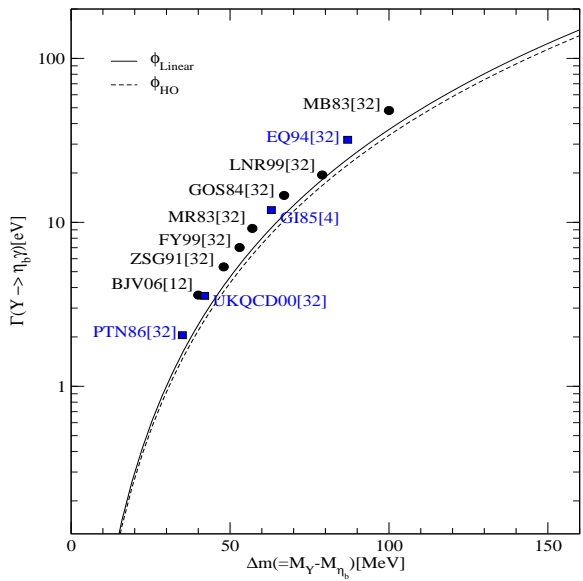}
\caption{\label{fig4}The dependence of $\Gamma(\Upsilon\to\eta_b\gamma)$ on
$\Delta m=M_\Upsilon - M_{\eta_b}$ compared with other theoretical
model calculations~\cite{UP}.}
\end{figure}
In Fig.~\ref{fig4}, we show the dependence of
$\Gamma(\Upsilon\to\eta_b\gamma)$ on $\Delta m$ compared with other
theoretical model calculations~\cite{UP}. As one can see from
Fig.~\ref{fig4}, our prediction for the dependence of 
$\Gamma(\Upsilon\to\eta_b\gamma)$ on $\Delta m$ is quite consitent with 
other theoretical predictions for various $\Delta m$~\cite{UP}.

\section{Summary and Discussion}
In this work, we investigated the weak decay constants and the
magnetic dipole $V\to P\gamma$ decays
of heavy-flavored mesons such as $(D,D^*,D_s,D^{*}_s,\eta_c, J/\psi)$ 
and $(B,B^*,B_s,B^*_s,\eta_b,\Upsilon)$ using the LFQM
constrained by the variational principle for the QCD-motivated effective
Hamiltonian. Our model parameters obtained 
from the variational principle uniquely determine the above nonperturbative 
quantities. This approach can establish the extent of applicability of our 
LFQM to wider ranging hadronic phenomena.

Our predictions of mass spectra and decay constants for heavy pseudoscalar 
and vector mesons are overall in good agreement with the available 
experimental data as well as other theoretical model calculations. 
Our numerical results
of the decay widths for $J/\psi\to\eta_c\gamma$ and $D^{*+}\to D^+\gamma$
fall within the experimental error bars. 
Our predictions for the branching ratios for the bottomed and 
bottomed-strange mesons are quite comparible with other theoretical model 
predictions.
For the radiative decay of the bottomonium, we find that the decay widths 
$\Gamma(\Upsilon\to\eta_b\gamma)$ is very sensitive to the value of 
$\Delta m=M_\Upsilon-M_{\eta_b}$. This sensitivity for the
bottomonium radiative decay may help to determine the mass 
of $\eta_b$ experimentally. 

Since the form fator $F_{VP}(q^2)$ of vector meson radiative decay 
$V\to P\gamma^*$ presented in this work is precisely analogous to the 
vector current form factor $g(q^2)$ in weak
decay of ground state pseudoscalar meson to ground state vector 
meson, the ability of our model to describe such decay is 
therefore relevant to the reliability of the model for
the weak decay. Consideration on such exclusive weak decays
in our LFQM is underway. Although our previous LFQM~\cite{CJ1,CJ2} and
this analyses did not include the heavy mesons comprising both 
$c$ and $b$ quarks such as $B_c$ and $B^*_c$, the extension of
our LFQM to these mesons will be explored in our future communication. 

\begin{acknowledgments}
This work was supported by a grant from Korea
Research Foundation under the contract KRF-2005-070-C00039. 
\end{acknowledgments}


\begin{thebibliography}{99}
\bibitem{CJ1}H.-M. Choi and C.-R. Ji,\Journal{\PRD}{59}{074015}{1999}.
\bibitem{CJ2}H.-M. Choi and C.-R. Ji,\Journal{\PLB}{460}{461}{1999}.
\bibitem{CJ3}H.-M. Choi, C.-R. Ji, and
L.S. Kisslinger, \Journal{\PRD}{65}{074032}{2002}.
\bibitem{GI} S.Godfrey and N. Isgur, \Journal{\PRD}{32}{189}{1985}.
\bibitem{Barik} N. Barik and P.C. Dash, \Journal{\PRD}{49}{299}{1994}.
\bibitem{Jaus96} W. Jaus, \Journal{\PRD}{53}{1349}{1996}.
\bibitem{EFG1} D. Ebert, R.N. Faustov, and V.O. Galkin,
\Journal{\PLB}{537}{241}{2002}.  
\bibitem{Cheng} H.-Y. Cheng {\em et al.}, \Journal{\PRD}{47}{1030}{1993}.
\bibitem{HQET} P. Colangelo, F. De Fazio, and G. Nardulli,
\Journal{\PLB}{316}{555}{1993}.
\bibitem{CBM} P. Singer and G.A. Miller, \Journal{\PRD}{33}{141}{1986};
\Journal{\PRD}{39}{825}{1989}.
\bibitem{Amu} J.F. Amundson, C.G. Boyd, I. Jenkins, M. Luke, A.V. Manohar,
J.L. Rosner, M.J. Savage and M.B. Wise, \Journal{\PLB}{296}{415}{1992}.
\bibitem{Nora}N. Brambilla, Y. Jia, and A. Vairo,
\Journal{\PRD}{73}{054005}{2006}.
\bibitem{EFG2} D. Ebert, R.N. Faustov, and V.O. Galkin,
\Journal{\PRD}{67}{014027}{2003}.
\bibitem{DA} H.-M. Choi and C.-R. Ji, \Journal{\PRD}{75}{034019}{2007}.
\bibitem{Data06} W.-M. Yao {\em et al.}(Particle Data Group), 
\Journal{\JPG}{33}{1}{2006}.
\bibitem{ZM} H.-M. Choi and C.-R. Ji, \Journal{\PRD}{72}{013004}{2005};
\Journal{\PRD}{58}{071901(R)}{1998};
B.L.G. Bakker, H.-M. Choi and C.-R.Ji,
\Journal{\PRD}{63}{074014}{2001}; \Journal{\PRD}{65}{116001}{2002};
\Journal{\PRD}{67}{113007}{2003}.
\bibitem{ALEPH} A. Heister {\em et al.}, ALEPH Collaboration,
\Journal{\PLB}{530}{56}{2002} and references therein.
\bibitem{Bec} D.Becirevic {\em et al.}, \Journal{\PRD}{60}{074501}{1999}.
\bibitem{Aubin} C. Aubin {\em et al.}(HPQCD Collaboration),
\Journal{\PRL}{95}{122002}{2005}.
\bibitem{Nar} S. Narison,\Journal{\PLB}{520}{115}{2001}.
\bibitem{Wang} G. Cvetic {\em et al.}, \Journal{\PLB}{596}{84}{2004};
G.-L. Wang, \Journal{\PLB}{633}{492}{2006}.
\bibitem{CG} S. Capstick and S. Godfrey, \Journal{\PRD}{41}{2856}{1990}.
\bibitem{Ebert} D. Ebert, R.N. Faustov and V.O. Galkin,
\Journal{\PLB}{635}{93}{2006}.
\bibitem{Gray} A. Gray {\em et al.}(HPQCD Collaboration),
\Journal{\PRL}{95}{212001}{2005}.
\bibitem{Hash} S. Hashimoto,
Int. J. Mod. Phys. A{\bf 20}, 5133(2005).
\bibitem{Jamin} M. Jamin and B.O. Lange,
\Journal{\PRD}{65}{056005}{2002}.
\bibitem{Cleo05}M. Artuso {\em et al.}, CLEO Collaboration,
\Journal{\PRL}{95}{251801}{2005}.
\bibitem{Cleo07}M. Artuso {\em et al.}, CLEO Collaboration, 
\Journal{\PRL}{99}{071802}{2007}.
\bibitem{Cleo_eta}K.W. Edwards {\em et al.}, CLEO Collaboration,
\Journal{\PRL}{86}{30}{2001}.
\bibitem{Belle_B}K. Ikado {\em et al.}, Belle Collaboration,
\Journal{\PRL}{97}{251802}{2006}.
\bibitem{GR} S. Godfrey and J. L. Rosner,
\Journal{\PRD}{64}{074011}{2001} and references therein.
\bibitem{UP} P. Moxhay and J.L. Rosner, \Journal{\PRD}{28}{1132}{1983};
R. McClary and N. Byers, \Journal{\PRD}{28}{1692}{1983};
H. Grotch, D.A. Owen, and K.J.Sebastian, \Journal{\PRD}{30}{1924}{1984};
F.J.Yndurain, hep-ph/9910399;X. Zhang, K.J. Sebatian, and H. Grotch,
\Journal{\PRD}{44}{1606}{1991}; T.A. L$\ddot{a}$hde, C.J. Nyf$\ddot{a}$lt,
and D.O. Riska, \Journal{\NPA}{645}{587}{1999};
Y.J.Ng, J. Pantaleone, and S.-H. H.  Tye, \Journal{\PRL}{55}{916}{1985};
UKQCD Collaboration, L. Marcantonio {\em et al.}, 
Nucl. Phys. B(Proc. Suppl.) {\bf 94}, 363(2001);
E. Eichten and C. Quigg, \Journal{\PRD}{49}{5845}{1994}.
\end{thebibliography}
\end{document}